\documentclass[12pt]{article}
\usepackage[T1]{fontenc}
\usepackage{tgtermes}
\usepackage{newtxmath}
\usepackage{natbib}
\usepackage{graphicx}
\usepackage{pifont}
\usepackage{multicol}
\usepackage{wrapfig}
\usepackage[utf8]{inputenc}
\usepackage[moderate]{savetrees}
\usepackage{ragged2e}
\usepackage{textgreek}
\usepackage{graphbox}
\usepackage{xcolor}

\usepackage{geometry}
\usepackage{color}
\definecolor{navy}{rgb}{0.0078, 0.4222,0.9059}

\usepackage[colorlinks=true]{hyperref}
\hypersetup{colorlinks, citecolor=navy, linkcolor=navy, urlcolor=navy}

\usepackage{enumitem,amssymb}
\newlist{thematic}{itemize}{8}
\setlist[thematic]{label=$\square$}

\usepackage{titlesec}
\titleformat*{\section}{\large\bfseries}
\titleformat*{\subsection}{\small\bfseries}

\newcommand{\hersc}{{\it Herschel}}

\newcommand{\spitz}{{\it Spitzer}}
\newcommand{\micron}{{\textmu m}}
\newcommand{\arcsec}{$^{\prime\prime}$}

\newcommand{\HI}{H{\sc i}}

\newcommand{\kappad}{$\kappa_{d}$}
\newcommand{\kappamu}{$\kappa_{500}$}

\begin{document}
\raggedright
\LARGE
Astro2020 Science White Paper \linebreak

{\bf Unleashing the Potential of Dust Emission as a Window onto Galaxy Evolution} \linebreak
\normalsize

\noindent \textbf{Thematic Areas:} \linebreak
$\bullet$ Galaxy Evolution \linebreak
$\bullet$ Star and Planet Formation \linebreak

\textbf{Principal Author:}

Name: Christopher Clark
\linebreak
Institution: Space Telescope Science Institute
\linebreak
Email: cclark@stsci.edu
\linebreak
Phone: (+1) 410-338-6813
\linebreak

\textbf{Co-authors:}\linebreak \justifying
Simone Bianchi, INAF OAA\\
Caroline Bot, ObAS\\
Viviana Casasola, INAF IRA\\
J\'er\'emy Chastenet, University of California (San Diego)\\
Asantha Cooray, University of California (Irvine)\\
Pieter De Vis, Cardiff University\\
Fr\'ed\'eric Galliano, AIM CEA/Saclay\\
Haley Gomez, Cardiff University\\
Karl Gordon, Space Telescope Science Institute\\
Benne Holwerda, University of Louisville\\
Julua Roman-Duval, Space Telescope Science Institute\\
Kate Rowlands, Johns Hopkins University\\
Sarah Sadavoy, Harvard-Smithsonian Centre for Astrophysics\\
Johannes Staguhn, Johns Hopkins University\\
Matthew Smith, Cardiff University\\
S\'ebastien Viaene, Ghent University\\
Thomas Williams, Cardiff University\\
\justifying


\pagebreak

\section{Introduction} \label{Section:Introduction}

Interstellar dust, the small grains of solid material that pervade interstellar space, provides us with a unique and powerful window onto the evolution of galaxies. Whilst dust only makes up \textless1\% of the InterStellar Medium (ISM) of galaxies by mass \citep{Remy-Ruyer2014A,Cortese2016C}, half of all the starlight emitted since the Big Bang has been absorbed by dust grains \citep{Driver2007D}. Over a third of the bolometric luminosity of Milky-Way-like galaxies is emitted by dust in the Mid-Infrared (MIR), Far-InfraRed (FIR), submillimeter, and millimeter parts of the spectrum \citep{Bianchi2018A}, whilst at high redshifts these wavelengths can completely dominate over the stellar emission \citep{Hughes1998C,Dunlop2017A}. And almost half the total mass of metals in the ISM of our Galaxy is locked up in dust grains \citep{Jenkins2009B}. 
However, our ability to truly exploit dust observations is stymied by {\color{red} dramatic, systematic uncertainties} in the behaviour of dust emission.

In Section~\ref{Section:Potential}, we describe science gains that studies of dust can yield. In Section~\ref{Section:Obstacles}, we describe the obstacles that currently harm our ability to reliably use dust emission as a tool for astronomy. In Section~\ref{Section:Solutions}, we describe the science program we propose in order to overcome these obstacles.

\section{The Potential of Dust} \label{Section:Potential}

Dust provides us with an unrivalled tool for probing the ISM of galaxies, and thereby uncovering how they evolve. Gas in galaxies can be directly observed through the emission from atomic gas, via the 21\,cm \HI\ line, and from molecular gas, typically via mm-range CO lines  -- however, both are extremely faint. The very largest extragalactic CO surveys are still only able to observe a few hundred targets, and mostly rely on measurements towards galaxies' centres, using assumed aperture corrections to infer total luminosity \citep{Saintonge2011A,Boselli2014B,Saintonge2017A}; whilst \HI\ is limited to redshifts of $z < 0.2$ (until SKA precursors enter full operation), which provides insufficient lookback time to trace evolution in the galaxy population \citep{Catinella2015B}.

Dust emission, however, has now been detected in over $10^{6}$ galaxies in FIR--mm surveys \citep{Schulz2017A,Maddox2018A}, with much of this due to the exceptional mapping abilities of \hersc\ and \spitz. Thanks to the relative brightness of dust emission, {\color{red} it will always be the case that orders of magnitude more galaxies will have their ISM detected via dust than via gas}. This advantage is compounded by the effect of negative $k$-correction, which means that the observed submm--mm brightness of galaxies of a given luminosity is almost unchanged throughout $1<z<6$, turning dust into one of the primary ways to study galaxies at high redshift \citep{Lutz2014A}. This means that dust provides the only means we have of uncovering the properties of the ISM in large statistical samples across the bulk of cosmic time -- {\color{red} assuming we can rely upon the dust properties we infer}. 

\begin{figure}
\centering
\includegraphics[width=0.375\textwidth,align=c]{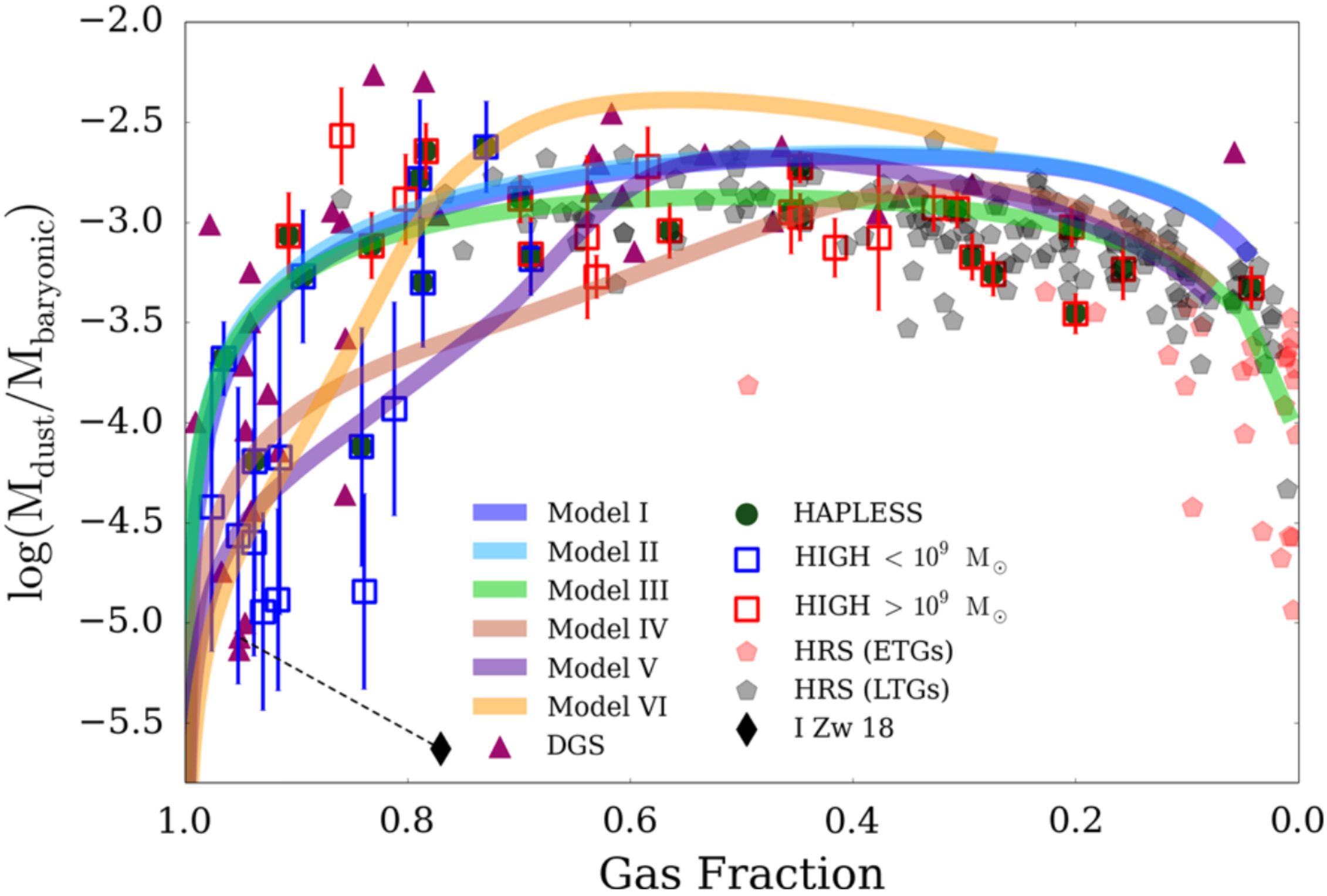}
\includegraphics[width=0.25\textwidth,align=c]{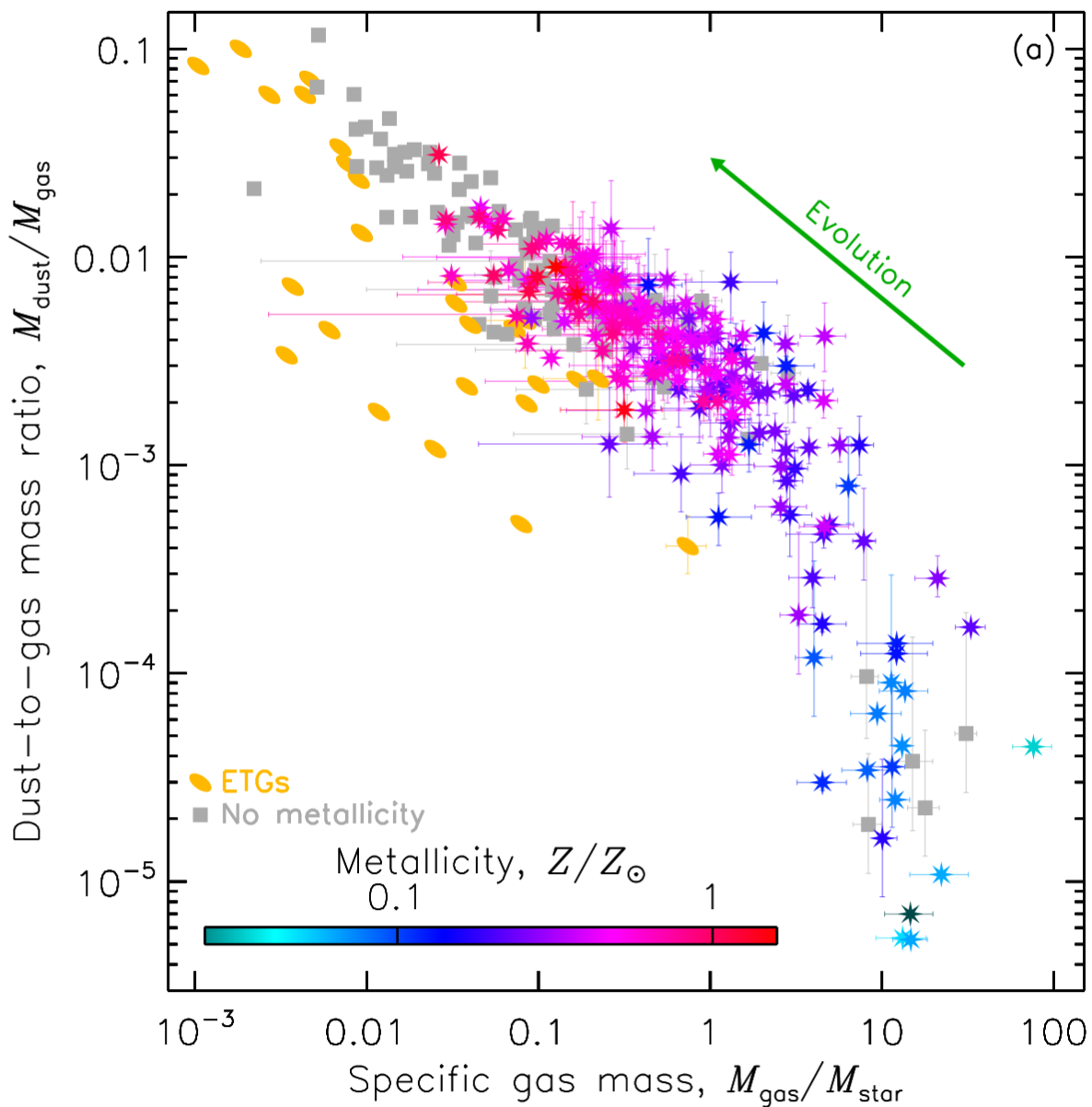}
\includegraphics[width=0.35\textwidth,align=c]{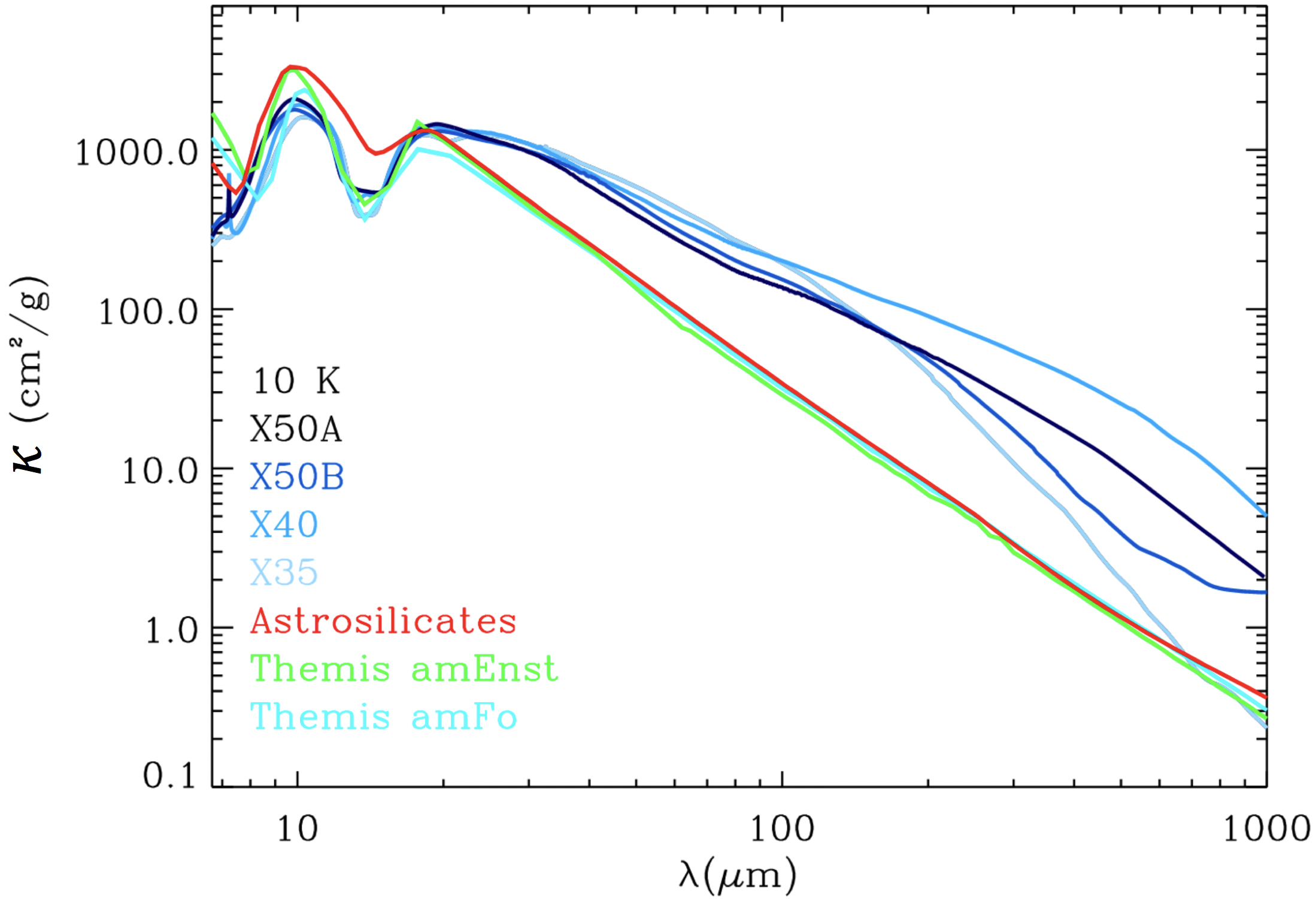}
\caption{{\it \bf Left:} Dust richness against gas fraction, with various evolutionary tracks overlaid; from \citet{DeVis2017B}. No one track can explain all observations, especially for  gas-rich systems. {\it \bf Centre:} Dust-to-gas ratio against gas richness, colour-coded by metallicity; from \citet{Galliano2018C}. Note the  `knee' at $\sim 0.3\,{\rm Z_{\odot}}$, where dust-to-gas appears to abruptly fall. {\it \bf Right:} \kappad\ against wavelength for the \citet{Draine2007A} model (`Astrosilicates'), the \citet{Jones2013C,Jones2016A} model (`Themis'), and the \citet{Demyk2017A} lab measurements of possible dust analogues (`X' samples); from \citet{Demyk2017A}. $\beta$ is the index of the power law slopes at $\lambda > 50$\,\micron\ (where $\kappa_{d} \propto \lambda^{-\beta}$).}
\label{Fig:Dust_Relations}
\end{figure}

Dust represents up to half the interstellar metal content of galaxies, making it a uniquely powerful tool for studying chemical evolution -- from young gas-dominated galaxies, to dust-rich intermediate systems, to galaxies exhausted of ISM \citep{CJRClark2015A,Schofield2016A,Kirkpatrick2017A}.  Dust studies have recently shown that there must be multiple, distinct evolutionary pathways taken by galaxies as they consume their ISM, even at the very earliest stages of their evolution (\citealp{DeVis2017B,DeVis2019A}; see left panel of Figure~\ref{Fig:Dust_Relations}).  And within galaxies, dust is now allowing us to dissect how they become enriched with heavy elements, disentangling the effects of star-formation histories, metal yields of different stellar deaths, galactic inflows/outflows, and environmental influences \citep{Cortese2016C,Roman-Duval2017B,Chiang2018A,Vilchez2019A}. Understanding chemical evolution in the local Universe is critical for understanding the dust budget crisis at intermediate redshift \citep{Rowlands2014B}, and the wide range of chemical evolutionary states seen at high redshift -- where some galaxies exhibit massive dust reservoirs at $z > 8$ \citep{Tamura2018A}, whilst others have very little dust, despite possessing abundant interstellar metals \citep{Matthee2017A}.

The dust-to-gas ratio is now an especially prominent tool for studying galaxy evolution, with a particular focus on the `critical metallicity' at which dust grain growth becomes significant \citep{Asano2013,Remy-Ruyer2014A,DeVis2019A}; see centre panel of Figure~\ref{Fig:Dust_Relations}. This has important implications for gas observations, because shielding by dust dictates how much CO can survive in the molecular phase, affecting its use as a tracer \citep{Sandstrom2013B,Amorin2016A,Accurso2017B}. Indeed, dust has now become a common proxy for estimating gas masses at higher redshifts, to exploit the speed with which dust observations can be performed, and sidestep the need to account for metallicity and excitation effects \citep{Eales2012A,Scoville2014B,Groves2015A}.

\section{The Obstacles} \label{Section:Obstacles}

Despite all the insights that dust can provide, there are major systematic uncertainties in our understanding of dust emission, which limit our ability exploit dust to its full potential.

\subsection{\kappad\ -- The Dust Mass Absorption Coefficient} \label{Subsection:Kappa_Obstacles}

\begin{wrapfigure}{l}{0.50\textwidth}
\centering
\includegraphics[width=0.50\textwidth]{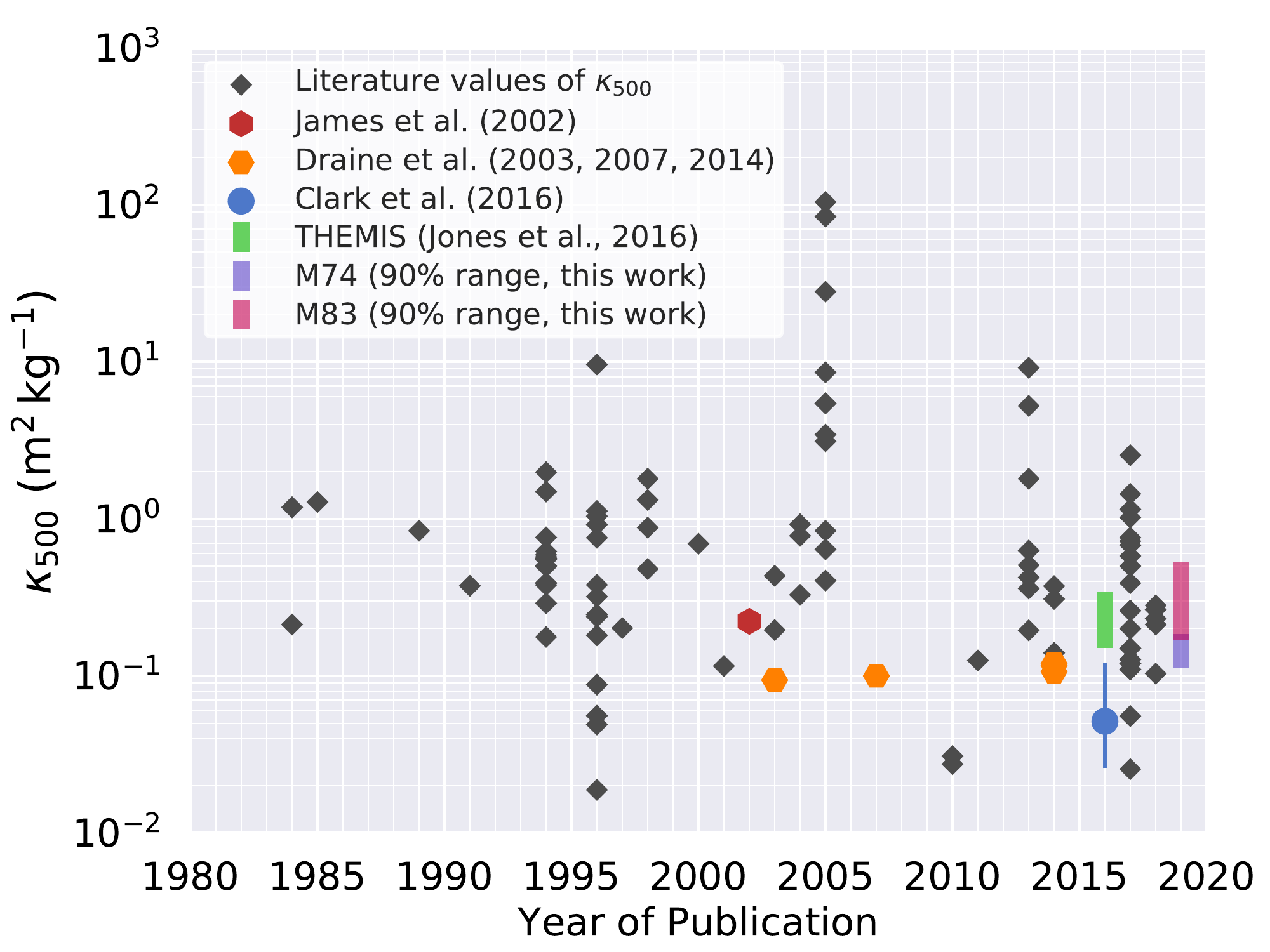}
\caption{Reported values of \kappamu\ (ie, \kappad\ at reference wavelength of 500\,\micron) against year of publication; from Clark et al. ({\it in prep.}). Includes values from theoretical models, empirical studies, and lab measurements.  Several values are highlighted with references, whilst a full list of references for all values can be found in Clark et al. (\it in prep.).}
\label{Fig:Year_vs_Kappa}
\end{wrapfigure}

The dust mass absorption coefficient, \kappad, is the conversion factor relied upon to infer physical dust masses from observations of dust emission (specifically, \kappad\  is a function of wavelength; see right panel of Figure~\ref{Fig:Dust_Relations}). All of the science described in Section~\ref{Section:Potential} used dust masses calculated with some assumed value of \kappad. However, the value of \kappad\ is notoriously uncertain -- reported values span {\color{red} \it several orders of magnitude} (see Figure~\ref{Fig:Year_vs_Kappa}), whilst even the most commonly-used values vary by a factor of $\sim 4$.

Worse yet, this uncertainty means it is standard practice to assume a constant value of \kappad\ -- whereas in reality, \kappad\ must vary both within, and between, galaxies. However, the degree to which these variations occur remains mostly unquantified. There are theoretical dust modelling frameworks -- such as the Draine et al. models \citep{Draine2007A,Draine2014A} and the THEMIS model \citep{Jones2013C,Jones2016A} -- that feature a mixture of grain types, with different emission properties \& abundances, the balance of which varies in different environments. However, these are primarily calibrated on Milky Way observations and laboratory analysis of likely dust analogues, limiting applicability to systems across the full range of metallicities, redshifts, etc. Meanwhile, empirical approaches to constraining \kappad\ are even more primitive, and typically depend upon assuming fixed values for the dust-to-metals ratio, emission-to-extinction ratio, or dust-to-gas ratio \citep{James2002,Planck2013XVII,CJRClark2016A,Bianchi2017A}.

It is common in the literature to see papers that describe a trend or phenomena seen in dust observations, and describe what physical process is believed to be responsible -- but then include a sentence saying that, alternatively, {\color{red} their findings could be all explained away by some factor X variation in \kappad} \citep{CJRClark2015A,Roman-Duval2017B,Chiang2018A,DeVis2019A}. Without confident knowledge of how \kappad\ varies, all of the advances described in Section~\ref{Section:Potential} are, to some extent, precarious.

The problem extends to the use of dust emission as a tracer for gas content at high redshifts. It is standard to use dust luminosity (often at 850\,\micron) as a direct proxy for molecular gas mass. Whilst this relationship is impressively robust for high-metallicity, massive, star-forming galaxies, extending it to broader galaxy populations will be much more complex \citep{Scoville2017C}, and will ultimately be most powerful when tied to the {\it mass} of dust, not just its luminosity.

\subsection{$\beta$ -- The Dust Emissivity Index} \label{Subsection:Beta_Obstacles}

As dust is not a perfect emitter, its spectrum is not a blackbody; rather, its emission in the Rayleigh-Jeans regime is modified according to an emissivity function, usually taken as $\sim\lambda^{-\beta}$. In short, $\beta$ dictates how \kappad\ varies with wavelength (illustrated in the right panel of Figure~\ref{Fig:Dust_Relations}); higher values of $\beta$ correspond to a steeper than a Rayleigh-Jeans tail. Lower values of $\beta$ are expected from metallic, crystalline, or carbonacous grains; whilst larger values of $\beta$ are expected from amorphous, silicate, or extremely cold grains \citep{Kohler2015A,Demyk2017A,Demyk2017B,Ysard2018A}. This means that $\beta$ is a direct indicator of dust properties.

\begin{figure}
\centering
\includegraphics[width=0.375\textwidth]{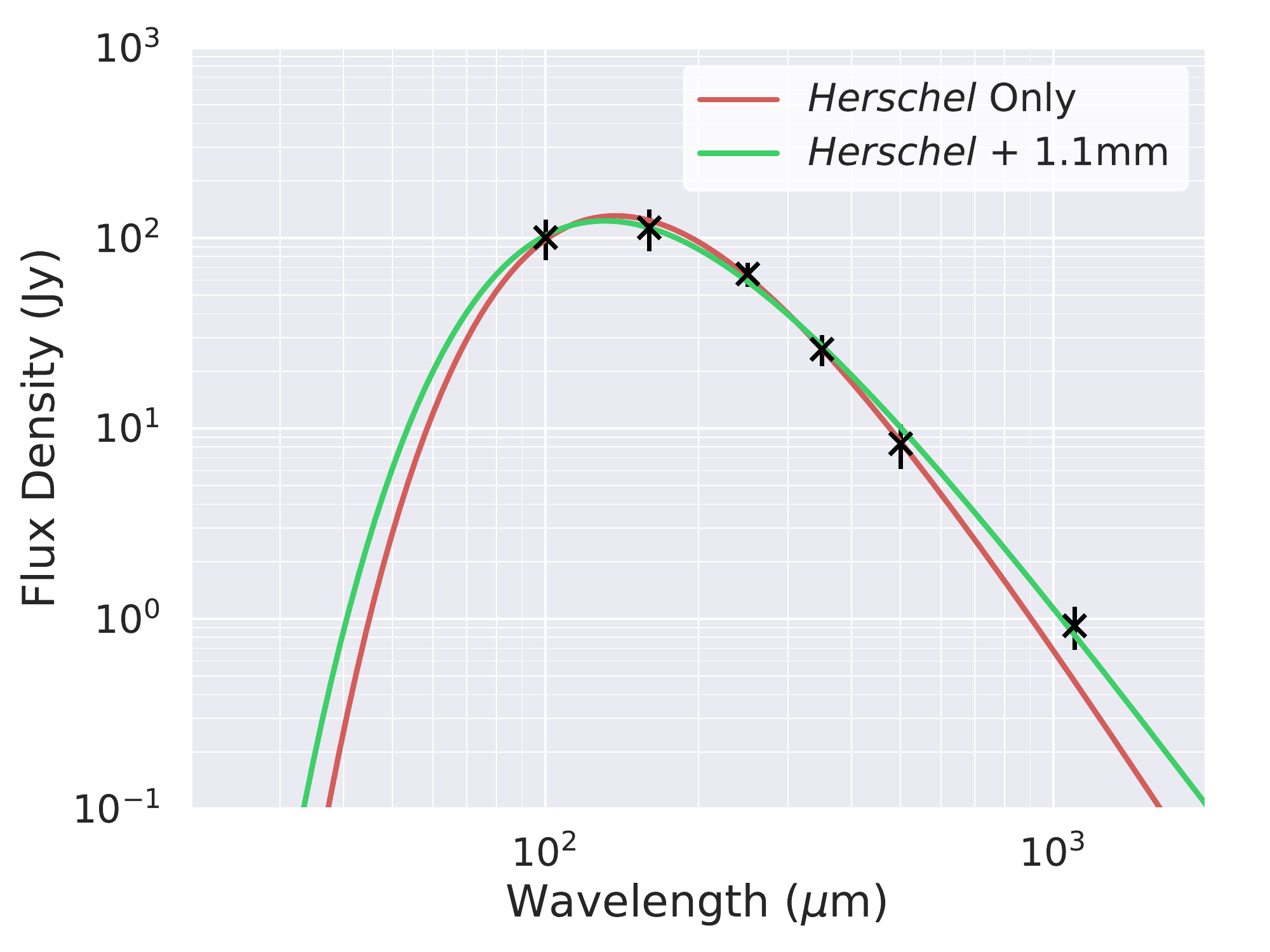}
\includegraphics[width=0.275\textwidth]{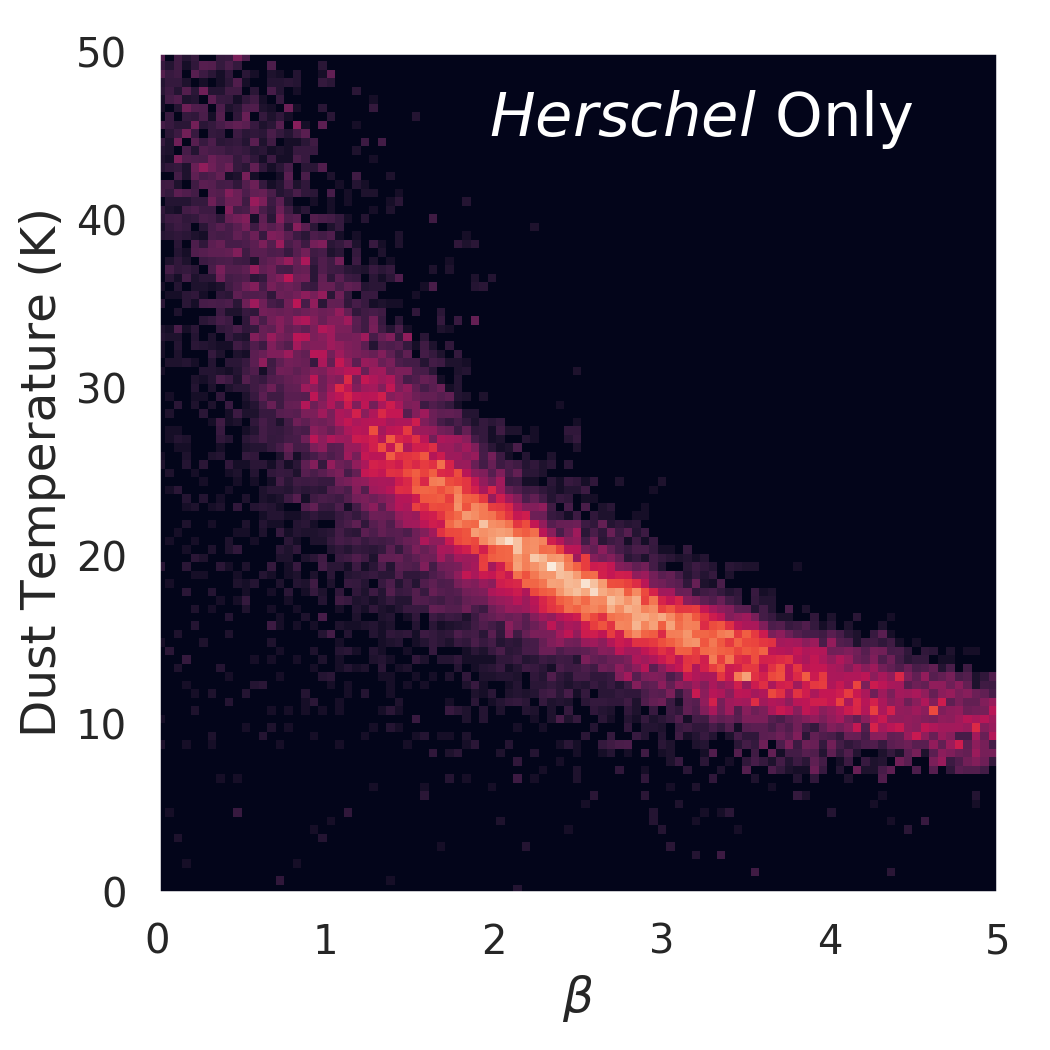}
\includegraphics[width=0.275\textwidth]{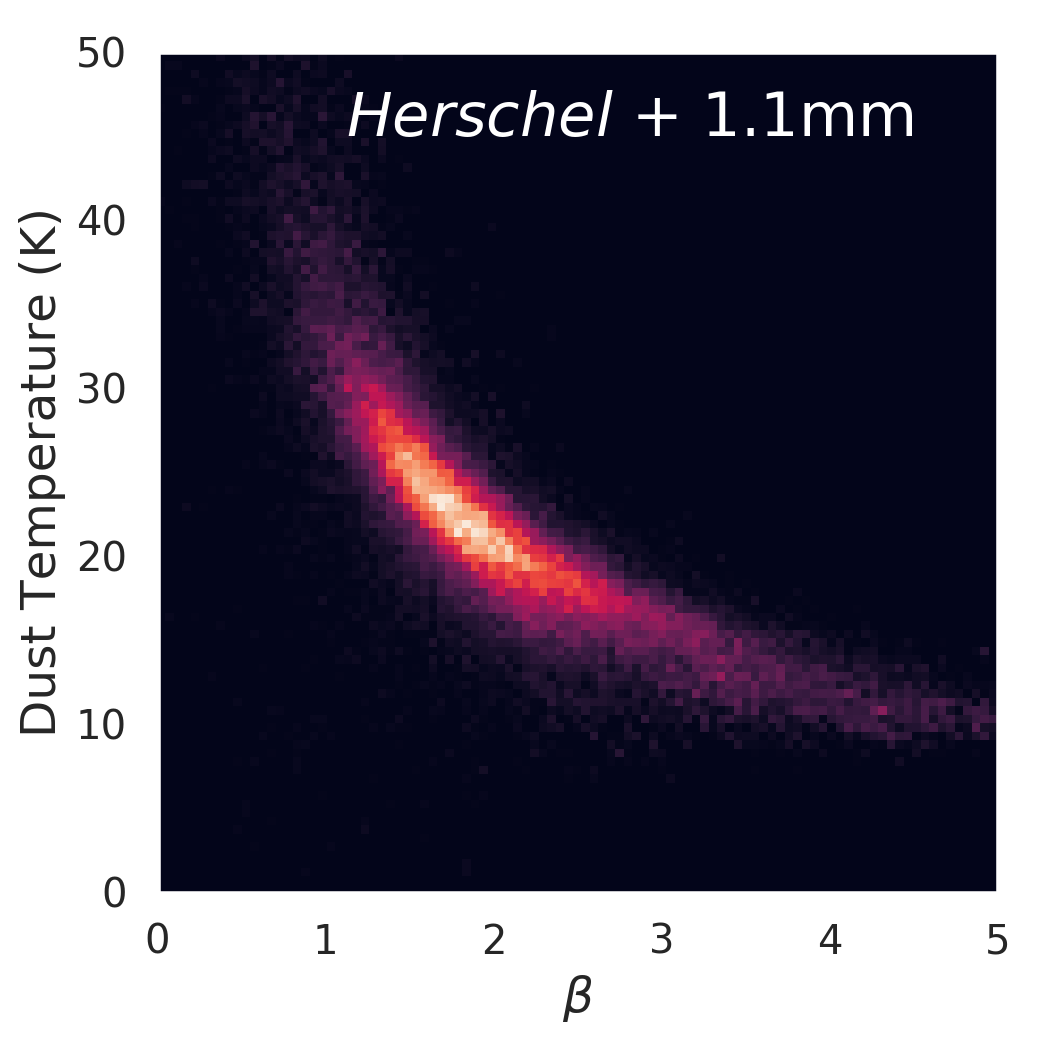}
\caption{{\it \bf Left:} SED model fitted to \hersc\ 100, 160, 250, 350, \& 500\,\micron\ fluxes only (red), and fitted to \hersc\ fluxes plus a 1.1\,mm flux (green). {\it \bf Centre \& Right:} Covariance plots of dust temperature with $\beta$ for the plotted SED models; the improvement due to the extra data is clear.}
\label{Fig:Beta_Improvement}
\end{figure}

However, in practice, $\beta$ is extremely hard to constrain. The primary reason for this is that when modelling the dust Spectral Energy Distribution (SED) of a galaxy, the value of $\beta$ is highly degenerate with the dust temperature \citep{Shetty2009A,Kelly2012B,Galliano2018C}; this is illustrated in Figure~\ref{Fig:Beta_Improvement}. This gives rise to an artificial anti-correlation between $\beta$ and temperature, thereby also limiting our ability to reliably interpret dust temperatures derived from SEDs. This in turn effects estimated dust masses, as the mass inferred from a dust SED scales as $\sim T^{\beta+4}$. 

Our difficulty in constraining $\beta$ is closely linked to the matter of the `submm excess' -- the observation of submm flux above what is predicted at submm--mm wavelengths. Whilst the effect appears strongest in dwarf galaxies (with up to 50\% excess emission at 500\,\micron; \citealp{Galliano2003A,Bot2010A,Gordon2014B,Grossi2015A}), it has also been observed in the low-density regions of spirals (up to 20\%; \citealp{Hunt2015A,Hermelo2016A}). Submm excess is equivalent to $\beta$ becoming shallower at longer wavelengths -- such strong variation with wavelength being an added complication to constraining $\beta$, and how it varies.

Additionally, the use of dust emission to estimate gas masses is typically done at wavelengths $\geq$\,850\,\micron; even at rest-frame, this corresponds to emission in the Rayleigh-Jeans regime, which is most sensitive to changes in $\beta$. 

\section{The Solutions} \label{Section:Solutions}

We envisage a program of study that will constrain the behaviour of \kappad\ and $\beta$. These investigations will not only uncover new facets of the properties of galaxies, and the dust they contain -- but also act as a multiplier,  enhancing the value of all research done via the window of dust emission. Here we will primarily focus on empirical avenues for progress.

The main limitation in tackling $\beta$ has been that {\color{red} the vast majority of galaxies observed in dust emission have no data available at wavelengths beyond the 500\,\micron\ \hersc\ point -- which is not far enough down the Rayleigh-Jeans tail to provide strong constraints} (see Figure~\ref{Fig:Beta_Improvement}). Even at 850\,\micron\ there is only moderate additional power to determine $\beta$ (\citealp{Sadavoy2013B,Rigby2018B}; Smith et al., {\it in prep.}; Lamperti et al., {\it in prep}); moreover, most galaxies detected at 850\,\micron\ are atypical (e.g. ultra-luminous infrared galaxies, submillimeter galaxies), and severe atmospheric noise makes mapping extended nearby galaxies problematic.

But over the coming decade, {\color{red} a new generation of 1--2\,mm large-area fast cameras}, such as MUSCAT \citep{Brien2018A}, NIKA2; \citep{Adam2018A}, and ToLTEC \citep{Bryan2018C}, are coming into service on large single-dish telescopes. Their mapping speeds (conservatively $10\,{\rm deg^{2}\,mJy^{-2}\,hr^{-1}}$ at 1.1\,mm) will allow them to detect detect $>10^{5}$ galaxies. For the facilities located at good sites (such as the LMT), it will be possible to map extended structure in nearby galaxies. They will achieve resolution as good as 5.5\arcsec\ -- corresponding to \textless\,200\,pc  for targets up to 7.5\,Mpc away (sufficient to resolve individual giant molecular clouds). This is a vital resolution regime, as it will allow us to trace the small physical scales, and therefore large dynamic ranges in density, over which ISM properties such as $\beta$ change \citep{MWLSmith2012B,Roman-Duval2017B,Williams2018A} -- with reduced mixing of physical conditions within the beam. These fast, large-area, mm-regime surveys will provide {\color{red} vital counterparts to the targeted observations performed by ALMA}; the exquisite resolution and sensitivity of ALMA has been the driving force behind ISM science in recent years, but also make it geared towards studying individual sources, of smaller angular size, in a limited number of bands.

Large surveys are of particular value in light of recent successes in fighting the $\beta$-temperature degeneracy with hierarchical Bayesian techniques \citep{Galliano2018B}, which rely on having large numbers of data points (ie, many pixels within a galaxy, or fluxes from many individual galaxies) to infer the true underlying parameter distributions. And, of course, the degeneracy can also be minimised by better constraining the dust temperature. This is best achieved in the MIR--FIR, where the dust SED peaks. The 30--90\,\micron\ regime in particular has been neglected (\spitz\ 70\,\micron\ had poor mapping speed, and the vast majority of extragalactic \hersc\ observations did not use the 70\micron\ band), and is ripe for exploitation; the SED shape here is sensitive to the presence of multiple dust components at a range of temperatures. {\color{red} Proposed missions such as Origins, GEP, or SPICA are needed} to obtain the large numbers of measurements in this range that are needed to employ the multi-phase multi-temperature dust models that are now under development \citep{Jones2017A,Ysard2018A}.

Synergistically, pinning down $\beta$ in different environments will aid in constraining variations in \kappad. Lab studies now show that $\beta$ should change for certain dust compositions, and that this will be more pronounced at colder temperatures \citep{Demyk2017A,Demyk2017B}. Thus characterising $\beta$ with long-wavelength observations will help inform the makeup of theoretical dust models -- and their \kappad\ values. This will be especially valuable at higher redshifts, where the nature and origin of dust is a key tracer of star formation and chemical evolution (see white paper `The Life Cycle of Dust'; Sadavoy et al.).

Understanding \kappad\ requires {\color{red} a `stalagmites and stalactites' approach} -- dust models advancing from one direction, empirical methods advancing from the other. On the empirical front, calibrating \kappad\ is ultimately a matter of knowing {\it a priori} how much dust there is in a system, and comparing that to the observed FIR--mm emission. The direct route to this is comparing the abundances of dust-forming metals in the neutral gas of the ISM, to their abundances in B stars that recently formed out of that ISM -- the difference being due to deplection of metals onto dust grains. Only \textless\,20 galaxies have had stellar abundances measured \citep{Kudritzki2012B,Bresolin2016B}, for only a few hundred stars in total -- and even this relied upon many hundreds of hours of time on 10\,m-class telescopes (eg, Keck, VLT). {\color{red} Meaningful advancement in this area will require the sensitivity of 30\,m-class telescopes}, and their ability to resolve giant stars beyond the current $\sim$5\,Mpc limit. In regards to abundances in the gas phase, we now have direct depletion measurements in some Local Group galaxies \citep{Roman-Duval2019A,Jenkins2017A}, but this requires UV spectroscopy, which will be lost after {\it Hubble} leaves service. We can also estimate total ISM abundances from nebular emission; 10\,m-class telescopes (such as the LBT) are starting to provide {\it direct} metallicities for large numbers of {\sc Hii}-regions in nearby galaxies \citep{Berg2015A}. And {\color{red} full metallicity {\it maps} for a few nearby galaxies}, determined via (less-reliable) strong-line relations, are now possible with IFU surveys -- a situation that will soon improve greatly with the SDSS-V Local Volume Mapper \citep{Kollmeier2017A}.

Once both neutral gas and stellar metallicity data is available, sampled at spatial resolution that matches observations of dust emission, we can determine the ISM dust-to-metals ratio directly, without having to assume a value of \kappad\ beforehand \citep{Davies2014A,Chiang2018A}, or rely upon depletion measurements that lack absolute normalisation \citep{Jenkins2009B}. Instead, we can simply take this direct dust-to-metals ratio, along with the gas-phase metal masses, and compare it to observations of dust emission -- and thereby finally directly compute robust empirical values for \kappad.


Finally we note that much of the science we describe here depends on facilities that are {\color{red} primarily closed-skies} (LMT, Keck, LBT), or {\color{red} oversubscribed by factors of 6-10} (VLT, ALMA). This raises the prospect of the vast majority of the US astronomical community being locked out of contributing to, and partaking in, vital progress in our field. Consideration should be given to how resources should be allocated to avoid such a situation -- especially in regard to submm-mm survey telescopes, where lack of recent national funding is costing the US community the access and leadership it used to enjoy.

\pagebreak
\begin{multicols}{2}
\def\ref@jnl#1{{\rmfamily #1}}%
\newcommand\aj{\ref@jnl{AJ}}%
\newcommand\araa{\ref@jnl{ARA\&A}}%
\newcommand\apj{\ref@jnl{ApJ}}%
\newcommand\apjl{\ref@jnl{ApJ}}%
\newcommand\apjs{\ref@jnl{ApJS}}%
\newcommand\ao{\ref@jnl{Appl.~Opt.}}%
\newcommand\apss{\ref@jnl{Ap\&SS}}%
\newcommand\aap{\ref@jnl{A\&A}}%
\newcommand\aapr{\ref@jnl{A\&A~Rev.}}%
\newcommand\aaps{\ref@jnl{A\&AS}}%
\newcommand\azh{\ref@jnl{AZh}}%
\newcommand\baas{\ref@jnl{BAAS}}%
\newcommand\jrasc{\ref@jnl{JRASC}}%
\newcommand\memras{\ref@jnl{MmRAS}}%
\newcommand\mnras{\ref@jnl{MNRAS}}%
\newcommand\pra{\ref@jnl{Phys.~Rev.~A}}%
\newcommand\prb{\ref@jnl{Phys.~Rev.~B}}%
\newcommand\prc{\ref@jnl{Phys.~Rev.~C}}%
\newcommand\prd{\ref@jnl{Phys.~Rev.~D}}%
\newcommand\pre{\ref@jnl{Phys.~Rev.~E}}%
\newcommand\prl{\ref@jnl{Phys.~Rev.~Lett.}}%
\newcommand\pasp{\ref@jnl{PASP}}%
\newcommand\pasj{\ref@jnl{PASJ}}%
\newcommand\qjras{\ref@jnl{QJRAS}}%
\newcommand\skytel{\ref@jnl{S\&T}}%
\newcommand\solphys{\ref@jnl{Sol.~Phys.}}%
\newcommand\sovast{\ref@jnl{Soviet~Ast.}}%
\newcommand\ssr{\ref@jnl{Space~Sci.~Rev.}}%
\newcommand\zap{\ref@jnl{ZAp}}%
\newcommand\nat{\ref@jnl{Nature}}%
\newcommand\iaucirc{\ref@jnl{IAU~Circ.}}%
\newcommand\aplett{\ref@jnl{Astrophys.~Lett.}}%
\newcommand\apspr{\ref@jnl{Astrophys.~Space~Phys.~Res.}}%
\newcommand\bain{\ref@jnl{Bull.~Astron.~Inst.~Netherlands}}%
\newcommand\fcp{\ref@jnl{Fund.~Cosmic~Phys.}}%
\newcommand\gca{\ref@jnl{Geochim.~Cosmochim.~Acta}}%
\newcommand\grl{\ref@jnl{Geophys.~Res.~Lett.}}%
\newcommand\jcp{\ref@jnl{J.~Chem.~Phys.}}%
\newcommand\jgr{\ref@jnl{J.~Geophys.~Res.}}%
\newcommand\jqsrt{\ref@jnl{J.~Quant.~Spec.~Radiat.~Transf.}}%
\newcommand\memsai{\ref@jnl{Mem.~Soc.~Astron.~Italiana}}%
\newcommand\nphysa{\ref@jnl{Nucl.~Phys.~A}}%
\newcommand\physrep{\ref@jnl{Phys.~Rep.}}%
\newcommand\physscr{\ref@jnl{Phys.~Scr}}%
\newcommand\planss{\ref@jnl{Planet.~Space~Sci.}}%
\newcommand\procspie{\ref@jnl{Proc.~SPIE}}%
\newcommand\jcap{\ref@jnl{JCAP}}%
\newcommand\eprint{\ref@jnl{arXiv e-print}}%

\bibliographystyle{aa}
\bibliography{ChrisBib}
\end{multicols}

\end{document}